\documentclass[pra,aps,twocolumn,showpacs]{revtex4} 
\usepackage{amsmath}
\usepackage{bm}              
\usepackage{graphicx}
\newcommand{\rl}{\rangle\!\langle}
\newcommand{\bi}{B_{i}}
\newcommand{\bid}{B_{i}^{\dag}}
\newcommand{\sumk}{\sum_{i}}
\begin{document}

\title{Second-order polaron resonances in self assembled quantum dots}

\author{Piotr Kaczmarkiewicz}
\author{Pawe{\l} Machnikowski}

\affiliation{Institute of Physics, Wroc{\l}aw University of
Technology, 50-370 Wroc{\l}aw, Poland}

\begin{abstract}%
We theoretically study the optical properties of an InAs/GaAs quantum dot (QD) near the area of the second-order resonance between an electron confined in the QD and two longitudinal optical phonons. We present the absorption spectra of an inhomogeneously broadened QD ensemble and show that the minimal model needed for an accurate description of such a system needs to account for 3-phonon states.
   We study also the influence of the QD height to width ratio on the optical properties of the polaron system. The dependence of the width of the resonance and the position of the second-order resonant feature on the height to width ratio is presented.
\end{abstract}


\maketitle   

\section{Introduction}
Interaction between confined carrier states and quanta of lattice vibrations is important in the description of both optical and transport properties of nanostructures. Coupling of discrete carrier states localized in a quantum dot (QD) with nearly dispersionless longitudinal optical (LO) phonons leads to formation of correlated electron-phonon states, i.e. polaron states \cite{hameau99,verzelen02a,hameau02}. Such coupling of electronic and phononic states results in the appearance of anticrossings whenever the energy of an unperturbed carrier state approaches the energy of one or multiple LO phonons \cite{hameau99,hameau02}. Particularly interesting is the resonance of second-order, that is, the resonance between a purely electronic state and a state with two LO phonons. Its description is nontrivial since there are no direct matrix elements coupling those states. In our recent works \cite{kaczmarkiewicz08,kaczmarkiewicz10}, we have shown that such a resonance can be accurately modeled using the Fr\"ohlich coupling Hamiltonian with standard (bulk) values of material parameters, contrary to earlier suggestions \cite{jacak02f}.

In this paper, we compare the calculated single QD polaron spectra, as well as intraband absorption spectra of inhomogeneously broadened QD ensembles obtained for different computational bases. We also study how the spectrum in the resonance area depends on the QD geometry (height to width ratio). 

The paper is organized as follows. First, in Sec. \ref{sec:model}, we present the model used for the description of an electron interacting with a phonon subsystem. Then, in Sec. \ref{sec:numres}, numerical results are presented. The effect of reducing the computational basis on the spectrum of the polaron is discussed, and the influence of the QD shape on the resonance parameters (the position and width of the resonance) is studied. The conclusions are presented in Sec. \ref{sec:conc}.


\section{Model}\label{sec:model}
We consider a symmetric, self-assembled InAs QD occupied by a single electron. The carrier confined in the parabolic potential of the QD interacts with the polarization field associated with LO phonons in the surrounding material (GaAs). The parabolic potential model remains qualitatively valid as long as the confinement supports a sufficient number of electronic shells (3 in our model, see below). Otherwise, a more general model, including the continuum states, would need to be used.

The Hamiltonian of the system can be expressed as
\begin{equation}\label{wzor:ham}
H=H_{0}+H_{\mathrm{int}}+H_{\mathrm{ph}},
\end{equation}
where $H_0$ is the Hamiltonian of an isotropic QD, $H_\mathrm{ph}$ is the free phonon term and $H_{\mathrm{int}}$ describes electron-phonon interaction.
The dynamics of an electron in the direction perpendicular to the QD plane  is restricted to the ground state of a harmonic oscillator with the out-of-plane confinement length $l_z=\sqrt{\hbar/(m^*\omega_z)}$, where $m^*=0.067\mathrm{m_e}$ is the effecitve mass of the electron and $\hbar \omega_z$ is the energy separation between eigenenergies of quantum harmonic oscillator in the $z$ direction. This is a very good approximation since the confinement in that direction is usually much stronger than the confinement in the $xy$ plane, that is, $\omega_0 \ll \omega_z$, where $\hbar \omega_0$ is the energy separation between the in-plane states.

The eigenfunctions of an electron in the two-dimensional parabolic potential are described by the Fock-Darwin states \cite{jacak98a}. The Hamiltonian describing the 2-dimensional states in an isotropic quantum dot,
\begin{equation*}\label{wzor:ham0}
H_{0}=\frac{{\bm{p}}^2}{2m^{*}} + \frac{1}{2}m^{*}\omega_0^2 r_{\bot}^2,
\end{equation*} 
can be expressed in the basis of Fock-Darwin states as
\begin{equation*} H_{0}=\sum_{nm}\epsilon_{nm}|nm\rl nm|,
\end{equation*}
where $n,m$ are radial and angular quantum numbers and the respective eigenvalues are
\begin{equation*} \epsilon_{nm}=\hbar\omega_{\mathrm{0}}(2n+|m|+1).
\end{equation*}

An important fact is that only the LO phonons from the vicinity of $\Gamma$ point are coupled to the electronic subsystem \cite{jacak03b}. Moreover LO phonons are nearly dispersionless for small wave vectors and it is possible to represent the phonons as a set of discrete collective modes rather than a continuum of states \cite{jacak03b,stauber00,obreschkow07}. 

The coupling between the confined electrons and LO phonons in the basis of Fock-Darwin states written in terms of collective phonon modes has the form \cite{kaczmarkiewicz10}.
\begin{eqnarray} H_{\mathrm{int}} & = & \sqrt{\frac{\hbar \Omega
e^2}{2l_B\varepsilon_0 \tilde \varepsilon}}
\sum_{nmn'm'}\sum_{\alpha}|nm\rl n'm'|  \nonumber\\ && \times
\gamma_{(nm)(n'm')\alpha} B_{m'-m,\alpha}  +\mathrm{H.c.}\nonumber,
\end{eqnarray}
where 
$\Omega$ is the frequency of LO phonons at the $\Gamma$ point, $l_0=\sqrt{\hbar/(m^*\omega_0)}$ is the in-plane characteristic length, $\epsilon_0$ is the vacuum permittivity, $\tilde{\varepsilon}= (1/{\varepsilon_\infty} -
1/{\varepsilon_\mathrm{s}})^{-1}=70.3$ is the effective dielectric constant, $\gamma_{(nm)(n'm')\alpha}$ are the coupling constants for the collective LO-phonon modes \cite{kaczmarkiewicz10}, and $B$ is collective phonon mode anihilation operator.

The last part of Eq. (\ref{wzor:ham}),
\begin{equation*} 
H_{\mathrm{ph}}=\hbar\Omega\sumk\bid\bi,
\end{equation*} describes free LO phonon term, where $i$ numbers all
different collective phonon modes included in the model.

After diagonalization of the Hamiltonian (\ref{wzor:ham}) one can easily calculate the intraband transition amplitudes between polaron ground state ($s$-like) and higher excited polaron states, which in the presented range of energies consist mostly of polaron $p$-like states.

As the intrinsic line width observed in the experiment \cite{hameau99} is much smaller than the ensemble inhomogeneity assumed in our numerical computations we do not include line broadening effects in our model. Formally, our modeling corresponds to the zero temperature situation.

\section{Numerical results}\label{sec:numres}
In our numerical modeling, we choose 3 lowest lying electronic shells (6 states). For such a number of electronic states, 14 different collective phonon modes need to be constructed. We restrict the computational basis to states with up to 4-phonons and show that a minimal model describing the second-order resonant polarons has to include states with up to 3-phonons. Although the presence of the 4-phonon states has a minor impact on the two-phonon resonance it can be important when considering higher order effects.


One of the methods of formally bringing electronic and phononic states to resonance is QD size adjustment. The energy separation between the $s$ and $p$ shell of confined electron states, $\hbar \omega_0$, decreases with growing dot size and for a certain size the resonance condition between the $s$-$p$ splitting and twice the LO phonon energy is met. Since it is the energy level splitting that is relevant for the spectral properties (rather than the QD size), it is convenient to use this quantity as an independent parameter in the model.

\begin{figure}[t]\label{fig:res}
  \includegraphics[scale=0.65]{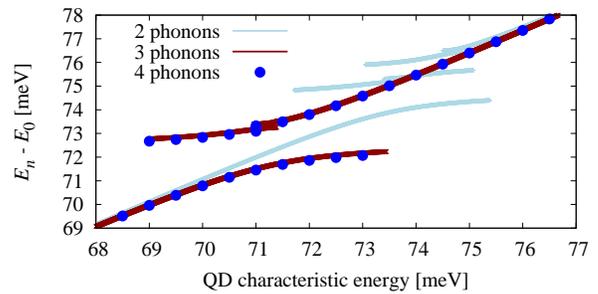}%
  \caption{\label{fig:1}%
Comparison between the polaron spectra for a single QD in the area of the second-order resonance obtained by diagonalization of the Hamiltonian for different truncations of basis states. The QD height to width ratio is $\lambda=0.2$.}
\label{onecolumnfigure}
\end{figure}
In Fig.~\ref{fig:1}, we present the energy spectrum for the case of a QD with a realistic shape ($\lambda=l_z/l_0=0.2$) for different numbers of phonon modes taken into account in the model. As can be seen, there is a big difference between the results obtained from models including 2- and 3-phonon states, whereas adding 4-phonon states does not produce any important changes in the presented area of the polaron spectrum.
If only states with up to two phonons are included an additional line appears in the area of the second-order resonance and the whole spectrum is shifted towards higher energies. When 3-phonon states are included this additional line in the middle of the resonance vanishes. This effect is due to uncoupling of a certain phonon mode which is strongly coupled in the absence of the 3-phonon states. A more detailed discussion based on a quasi-degenerate perturbation theory has been presented in our previous paper \cite{kaczmarkiewicz10}.
\begin{figure}[t]\label{fig:res-abs}
  \includegraphics[scale=0.65]{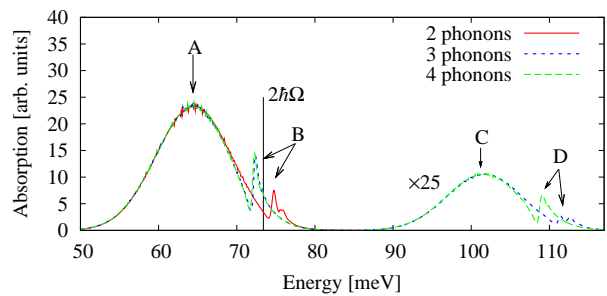}%
  \caption[]{\label{fig:2}%
   Comparison of absorption spectra for different truncation of computational basis. Feature A -- the main absorption peak, B -- the two-phonon resonance, C -- phonon replica of main absorption peak, D - phonon replica of the second-order resonance. The QD height to width ratio $\lambda=0$, mean transition energy $\overline{\epsilon}=63~\mathrm{meV}$, and standard deviation $\sigma=5~\mathrm{meV}$.}
\label{onecolumnfigure}
\end{figure}

For a given distribution of QD characteristic energies $\hbar \omega_0^{(i)}$, the absorption spectrum of an inhomogeneously broadened QD ensemble can be calculated. The distribution of QD characteristic energies is assumed to be described by a
Gaussian function 
\begin{equation*}
\label{eq:gauss} 
f_{\overline{\epsilon},\sigma}(\hbar \omega_0)=
\frac{1}{\sigma \sqrt{2\pi}} e^{-\frac{(\hbar \omega_0-\overline{\epsilon})^2}{2
\sigma^2}}, 
\end{equation*} 
where $\overline{\epsilon}=\hbar \overline{\omega_0}$
and $\sigma$ are the mean transition energy and its standard
deviation, respectively.

The absorption spectra of an inhomogeneously broadened QD ensemble calculated using models with different number of phonons are presented on Fig.~\ref{fig:2}. Again, in the second-order resonance area (B), only the difference between models with up to 2 and 3 phonons is particularly large. After including 3-phonon states the position of the resonance shifts towards lower energies and the resonant feature (B) consists of a sharp peak only. Influence of the additional 4-phonon states on the second-order resonance area is negligible, but it is quite large if one considers the phonon replica of the second-order resonance, that is, resonance between 1-phonon and 3-phonon states (feature D). The presence of 4-phonon states has similar effect on this area as including the 3-phonon states had on the area of the second-order resonance.

\begin{figure}[th]\label{fig:lambda_infl}
  \includegraphics[scale=0.65]{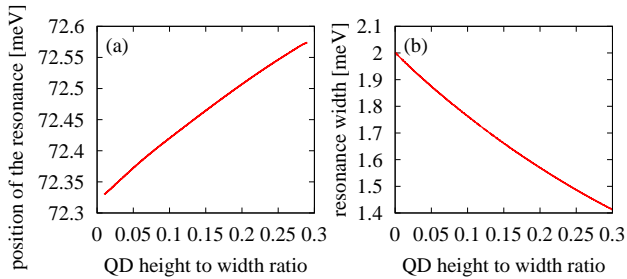}%
  \caption{\label{fig:3}
     Influence of QD height to width ratio on the position of the resonant feature (a) and on the width of the resonance (b).}
\label{onecolumnfigure}
\end{figure}

The influence of the QD shape (height to width ratio $\lambda$) on the system spectrum in the resonance area is shown in Fig.~\ref{fig:3}. In a single-dot polaron spectrum it shifts the resonance and changes its width. For QDs with a larger value of $\lambda$, the electron-LO-phonon coupling constants have smaller values, thus the width of the resonance decreases. As it turns out, changing this shape parameter does not produce strong shifts in the position of the second-order resonant feature. 
Since the position of the resonant feature is a nearly linear function of the QD height to width ratio the absorption spectrum of an ensemble of QDs differing not only in sizes but also in heights can be characterized by average of $\lambda$ parameter.
\section{Conclusions}\label{sec:conc}
In this paper, we have theoretically modeled the second-order resonance between $p$-shell electron states confined in a quantum dot and $s$-shell state with 2 LO phonons. We have studied the effect of restricting the computational basis on the absorption spectrum of an inhomogeneously broadened QD ensemble. The presented data confirm that including 3-phonon states is necessary in modeling the second-order resonant polarons. When considering higher order effects a theoretical model has to account also for 4-phonon states. 

We have also studied the dependence of the polaron spectrum on the QD geometry.
We have shown that both the position of the resonance and its width depend on the QD height to width ratio. Due to stronger interaction between electrons and phonons in the case of flat dots, the width of the resonance increases with decreasing height to width ratio. The character of this dependence as well as the dependence of the position of the resonance on the QD height to width ratio is nearly linear. 

The presented data confirm that the resonant feature is clearly visible in the absorption spectra and the system can be modeled with standard material parameters and computational techniques, which provides a way towards modeling of even more complicated electron-phonon systems.


%
%

\end{document}